\documentclass[%
pre
,twocolumn
,letterpaper
,showkeys%
,showpacs%
,floatfix%
,amsmath]{revtex4}

\usepackage[T1]{fontenc}

\usepackage{ifpdf}
\ifpdf
\usepackage{txfonts}
\usepackage{microtype}
\usepackage[colorlinks,linkcolor=blue,citecolor=blue,urlcolor=blue]{hyperref}
\fi

\usepackage{graphicx}

\begin{document}


\title{A spatially explicit model for tropical tree diversity patterns}

\author{Zolt\'an N\'eda}
\affiliation{%
Babe\c{s}--Bolyai University, Faculty of Physics, \\
M.~Kog\u{a}lniceanu 1, 400084 Cluj-Napoca, Romania
}

\author{Szabolcs Horv\'at}
\affiliation{%
University of Bergen, Department of Physics and Technology, \\
All\'egaten 55, N-5007, Bergen, Norway
}

\author{Hajnalka M\'aria T\'oh\'ati}
\affiliation{%
University of Szeged, Dept.~of Physics, \\
D\'om t\'er 9, H-6720 Szeged, Hungary
}

\author{Aranka Derzsi}
\affiliation{%
Hungarian Academy of Sciences, Research Institute for Solid State Physics and Optics, \\
H-1525 Budapest, Hungary
}

\author{Adalbert Balogh}
\affiliation{%
EMTE--Sapientia University, Dept.~of Zoology, \\
Tg.~Mure\c{s}/Corunca, Romania
}

\date{\today}

\begin{abstract}
A complex two-parameter model resembling the classical voter model is introduced to describe macroecological properties of tropical tree communities. Monte-Carlo type computer simulations are performed on the model, investigating species abundances and the spatial distribution of individuals and species. Simulation results are critically compared with the experimental data obtained from a tree census on a 50 hectares area of the Barro Colorado Island (BCI), Panama.
The model parameters are optimized for reproducing quantitatively the experimental results from the BCI dataset.  
\end{abstract}


\maketitle

\section{Introduction}
On long time scales---evolutionary time---the particular characteristics of species evolve widely from birth to extinction, while throughout ecological time most species persist sometimes for decades or even centuries \cite{hubbell,brown}. Their organization in complex networks of local communities and extended metacommunities is subject to abundant variations in terms of relative distributions. Usually, species maintain themselves with constant abundance but, from time-to-time, processes such as invasion or succession produce fast regrouping throughout a particular site or region \cite{tokeshi}.

Macroecology \cite{brown,pielou,bell_2001} studies the relationships between organisms and their environment at large spatial scales in order to characterize and explain universal statistical patterns of abundance, distribution and diversity \cite{may_1975,harte_scaling_2000,harte_self-similarity_1999}. A variety of models and methods from statistical physics are appropriate for the study of ecosystem dynamics, in which the main ``entities'' are either many individual organisms within populations, or many species within local, regional or continental communities \cite{dewdney_dynamical_2000,chave_neutral_2004,enquist_modeling_2002,norris_neutral_2003,chave_comparing_2002}.

The remarkable regularities in patterns of data on how species originates, persist, assemble in groups, and eventually go extinct, suggest the existence of general mechanisms from which the biodiversity and the structure of ecological communities originate \cite{brown,hubbell,tokeshi}. The difficulties in explaining the biological diversity of such systems originate from the very different spatial and temporal scales, starting with the evolution and biogeographic distribution of species and ending with individual births and deaths in local communities. Present approaches for describing the dynamics of population genetics and ecology consider either the importance of the genetic fitness or the influence of random events governing birth, death and migration phenomena \cite{hubbell}. There are thus two main, but conflicting perspectives on the nature of ecological communities: the niche-assembly and the dispersal-assembly perspective \cite{brown,hubbell}. In the present work we intend to use this second approach, constructing a simple spatially explicit stochastic model for describing the relative species abundance and the spatial diversity patterns of trees in a tropical forest \cite{condit_beta-diversity_2002,volkov_neutral_2004,volkov_patterns_2007}.

\section{Neutral models}

Nowadays scientist argue that ecological systems of similar species that compete with each-other only for the limited amount of natural resources can be successfully approached by means of neutral models. One of the first and most widely used neutral models is that of Stephen Hubbell \cite{hubbell,volkov_neutral_2003},
who assumed in his theory that within groups of ecologically similar species, individuals fill the landscapes up to a point of saturation and the dynamics of all these species is governed by the same birth-, 
death\nobreakdash-, migration- and mutation-rates. This simple neutrality principle allows us to study a wide variety of systems by means of stochastic computer simulations and to use the methods of statistical mechanics to get analytical results \cite{alonso_merits_2006,azaele_linear_2006,etienne_neutral_2007,pigolotti_stochastic_2004,volkov_organization_2004}.

Tropical forests \cite{condit_beta-diversity_2002,volkov_neutral_2004,volkov_patterns_2007,condit_census} contain a wider variety of tree species compared to northern coniferous forests (2 orders of magnitude larger), and it is believed that such systems are much closer to the neutrality assumption. Since the number of mature individual trees in equal-area samples of each forest type is almost the same, the increase in the abundance of a particular species needs to be compensated by a decrease in another species's population. 

In order to develop an appropriate model, one needs to take into account the whole set of factors influencing community composition: births, deaths, immigration and (on longer time scales) speciation. These processes interact in a complex manner and produce the commonly observed empirical patterns of diversity and distribution of abundances. The model should prescribe, for each generation, a method for choosing the species identity of the individuals that die and those that will occupy (by birth or immigration) the vacancies created. A simple method which describes fairly well the above mentioned processes is the one developed by Hubbell and it is called ``zero-sum ecological drift'' \cite{hubbell,volkov_neutral_2003}. This method implies that the replacing species are drawn at random from the existing community of species. The ``ecological drift'' is connected to the random replacement process and assumes that all individuals, regardless of species, have equal probabilities of giving birth, dying, immigrating or acquiring a mutation to generate a speciation event. The ecological drift does not imply an equal chance for each species to fill a given vacancy. Obviously, the greater probability of being drawn into a vacancy belongs to more abundant species. A remarkable observation is that individuals are equal, but species, as collective entities, are not. In Hubbell's model \cite{hubbell,volkov_neutral_2003}  the ecological drift, with no additional mechanism is enough to produce the patterns of species abundance and diversity observed in nature. In particular, the community of tropical trees is very well described by these neutral models and neutrality will be also the basis of the approach considered in this work. Contrary to earlier modeling efforts \cite{hubbell,volkov_neutral_2003,volkov_density_2005}, here we will continue the novel idea introduced by Zillio~et~al.\ \cite{zillio_spatial_2005}, and a spatially explicit model resembling the classical voter model will be studied. In our approach we will consider, however, a more complex, two-parameter version, which is not suitable for an analytical study. We will argue why it is important to make the model considered in \cite{zillio_spatial_2005} more complex, and we will determine the best parameters for optimally reproducing the measured macroecological patterns.

\section{Relevant macroecological measures}

In order to describe the statistics of species sizes in meta- or local ecological communities, the main measure which is usually considered is the Relative Species Abundances (RSA) and the Rank-Abundance distribution. Provided our data is spatially accurate, and we know the location of each individual, we can also investigate the species-area scaling and the auto-correlation function for the individuals of a given species. We will discuss briefly these measures here. 

\subsection{Relative species abundance distribution (RSA)}

\begin{figure}[ht]
  \begin{center}
    \includegraphics{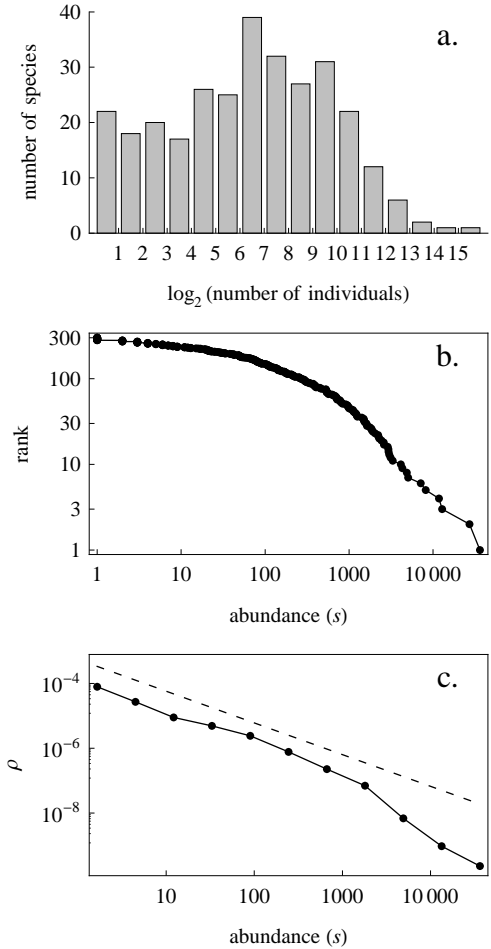}
  \end{center}
  \caption{RSA for the 50 ha Barro Colorado Island tropical tree census. Three different ways of presenting RSA: (a) the Preston plot, (b) the rank-abundance plot, and (c) the probability density function (the dashed line indicates 
a power-law with exponent $-1$).}
  \label{fig:bci_rsa}
\end{figure}

The relative species abundance distribution is introduced for characterizing the frequency of species with a given abundance \cite{volkov_patterns_2007,preston_canonical_1962}. Three different types of plots are generally used (Fig.~\ref{fig:bci_rsa}) for representing species-abundance distribution. Historically the first, and thus also the most widespread, representation of species abundances is due to Preston \cite{preston_canonical_1962}, who considered exponentially increasing abundance intervals (i.e.\ $a^i$, $i=1,2,\ldots$ with fixed $a$), and plotted the number of species found within these intervals as a function of $i$.  Preston's plot is motivated by the fact that abundances can vary in wide limits and there are relatively small numbers of abundant species. By considering fixed length abundance intervals one would get large statistical fluctuations on the tail of the curve. For widely different communities, the Preston plot has a gaussian-like shape. As an example for such a representation, on Fig.~\ref{fig:bci_rsa}a we illustrate the shape of RSA for the Barro Colorado tropical forest tree census. In this case however, the curve does not show a clear gaussian shape.  

A second way of representing the species abundance distribution is arranging the species in decreasing order by their abundances and plotting the species's rank versus its abundance (rank-abundance plot) on a log-log scale \cite{chave_neutral_2004,norris_neutral_2003,chave_comparing_2002}. This type of representation is inspired by several abundance studies in sociology and economics, leading to the very general Pareto-Zipf distribution. For the Barro Colorado tropical tree census the rank-abundance plot is illustrated on Fig.~\ref{fig:bci_rsa}b.

The third and mathematically most rigorous way of representing the species-abundance distribution is plotting the mathematical distribution function $\rho(s)$, i.e.\ the probability density for finding a species with a given $s$ abundance.  For most of the neutral-like communities this distribution function has a tilted J shape on a log-log scale. As an example, on Fig.~\ref{fig:bci_rsa}c we plot this probability density for the Barro Colorado tropical tree census. It is worth mentioning that $\rho(s)$ can be derived from the Preston plot by dividing the number of individuals in each interval with the length of the interval, and plotting on log-log scale this quantity versus the mean abundance in the given interval.  It is also easy to realize that the rank-abundance plot is related to the cumulative distribution function corresponding to the $\rho(s)$.

\subsection{Species-area relationship (SAR)}

\begin{figure}[ht]
  \begin{center}
    \includegraphics{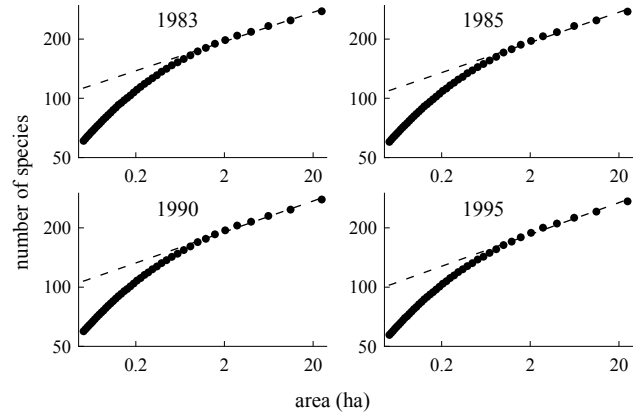}
  \end{center}
  \caption{Species-area scaling for different BCI census data. The continuous line suggests the power-law fit in the limit of large areas.  Scaling exponents in this limit are in the range 0.15--0.16 for the different census years. 
Both scales are logarithmic.}
  \label{fig:bci_sar}
\end{figure}

Generally, the number of detected species does not scale linearly by increasing the size of the sampled territory. In order to characterize this dependence, the species-area relationship is studied. This is done by considering larger and larger territories, and counting the number of species present within these areas \cite{harte_scaling_2000,harte_self-similarity_1999,zillio_spatial_2005}. For a better statistic, sometimes an average species number is calculated on several territories with similar areas. The dependence of the average species number on the size of the sampling area is usually a power-law, with exponent in the range 0.15--0.5. As usual, the presence of the scaling is visually illustrated by using a log-log scale like in (Fig.~\ref{fig:bci_sar}). 

\subsection{The  auto-correlation function}

In case the individuals are restricted to a given spatial position, like trees for example, one can characterize the spatial distribution of the individuals from a given species using an auto-correlation function. This function is defined as the usual auto-correlation function in statistical physics, and characterizes the probability density that for a given individual an individual from the same species exists at a distance $r$. The way we construct the $C(r)$ auto-correlation function is the following. First a check-board type uniform mesh is considered, and the territory is divided into small, square-like domains (labeled by coordinates $i,j$) and the number of individuals from the considered species is determined in each domain $(N_{i,j})$. The auto-correlation function for a relative coordinate $p,q$  ($C_{p,q}$) is calculated as
\newcommand{\avgN}{\ensuremath{\langle N \rangle}}
\begin{equation}
  C_{p,q} = \bigl<(N_{i,j} - \avgN) (N_{i+p,j+q} - \avgN) \bigr>_{i,j}
\end{equation}
Here \avgN{} denotes the average number of individuals from the considered species in the constructed domains: $\avgN = \langle N_{i,j} \rangle_{i,j}$. Since there is no reason to assume that the distribution of the individuals is non-isotropic, we can calculate the average of the $C_{p,q}$ values for all $p,q$ values that are inside a ring with radius $r$ and width $\Delta r$ ($r \le \sqrt{p^2 + q^2} \le r + \Delta r$), considering a reasonable small $\Delta r$ value.
\begin{equation}
  C(r) = \bigl< C_{p,q} \{ r \le \sqrt{p^2 + q^2} \le r + \Delta r \} \bigr>_{p,q}
\end{equation}

\begin{figure}[ht]
  \begin{center}
    \includegraphics{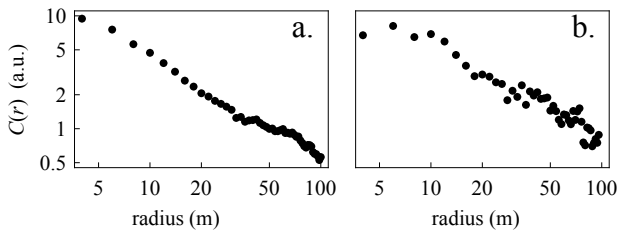}
  \end{center}
  \caption{The $C(r)$ spatial correlation function for the species that is most abundant among trees with a diameter larger than 10 cm (\emph{Trichilia	tuberculata}). Both scales are logarithmic. Figure~a.\ shows the results considering all individuals and Figure~b.\ is obtained for trees with diameter larger than 10 cm.}
  \label{fig:bci_corr}
\end{figure}

\section{Experimental data: the Barro Colorado Island dataset}

The experimental results used for comparison are from a detailed 50 hectares tropical forest tree census in Barro Colorado Island (BCI), realized by the Smithsonian Tropical Research Institute, Center for Tropical Forest Science (CTFS) \cite{condit_census}. BCI is located in the Atlantic watershed of the Gatun Lake (Panama) and was declared a biological reserve in 1923. It has been administrated by the Smithsonian Tropical Research Institute since 1946. From the viewpoint of ecological studies, this island is ideal, because it is covered with a rain forest that is still unperturbed by human beings. The flora and fauna of Barro Colorado Island have been studied extensively and inventories have reported 1369 plant species, 93 mammal species (including bats), 366 avian species (including migratory), and 90 species of amphibians and reptiles. The tropical tree census was performed only on a small part of this huge island, precisely on a $1000 \times 500$ meters (50 hectares) area. The first census was completed in 1982, revealing a total of approximately 240{,}000 stems of 303 species of trees and shrubs more than 1 cm in diameter at breast height. This CTFS program has an important advantage, because in each census, all free-standing woody stems at least 10 mm diameter at breast height are identified, tagged, and mapped, and hence accurate statistics can be made. Data is publicly available for several years: 1982, 1985, 1990 and 1995 \cite{bci_data}.

\begin{figure}[ht]
  \begin{center}
    \includegraphics{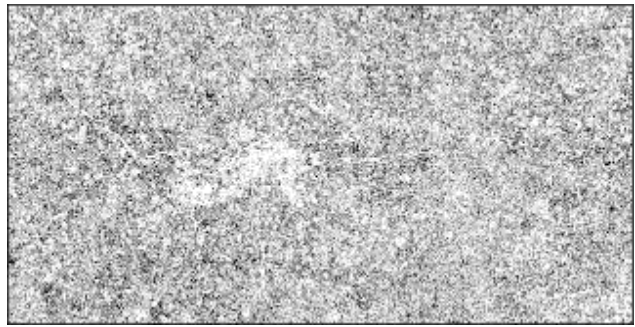}
  \end{center}
  \caption{Density of trees in the 1990 BCI census. The grayscale code indicates the density, darker squares indicate higher tree densities.}
  \label{fig:bci_dens}
\end{figure}

\begin{figure}[ht]
  \begin{center}
    \includegraphics[width=240pt]{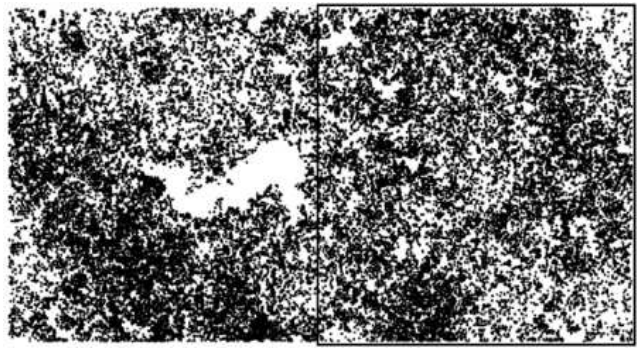}
  \end{center}
  \caption{Spatial distribution of the most abundant species: \emph{Hybanthus prunifolius}. The square on the left shows the region in which the relevant macroecological measures were computed.}
  \label{fig:hybanthus_dens}
\end{figure}

From the BCI data file it results that the studied area embodies 316 tree species, containing in total more than 320{,}000 individuals \cite{condit_census}. The density of individuals for the 1995 census is illustrated on Fig.~\ref{fig:bci_dens} with a gray-scale code (darker region corresponds to higher density). In the western part of the studied area lies a relatively large swampland (lighter region in the left side of Fig.~\ref{fig:hybanthus_dens}, i.e.\ less trees). This area is ecologically quite different from the other parts (see for example the spatial distribution of the most abundance species: \emph{Hybanthus prunifolius} (Fig.~\ref{fig:hybanthus_dens}). Therefore in determining the relevant macroecological measures only trees located in the eastern square-like part of this region ($500 \times 500$ meters) were considered.

The relative species abundance distribution for the census from 1990 is illustrated in Fig.~\ref{fig:bci_rsa}. 
The rank-abundance plot and the probability density curves suggest a scaling with an obvious cutoff. The scaling exponent is close to $-1$ ($-0.98$). Data for the other census years gives very similar results. 

\begin{figure}[ht]
  \begin{center}
    \includegraphics[width=240pt]{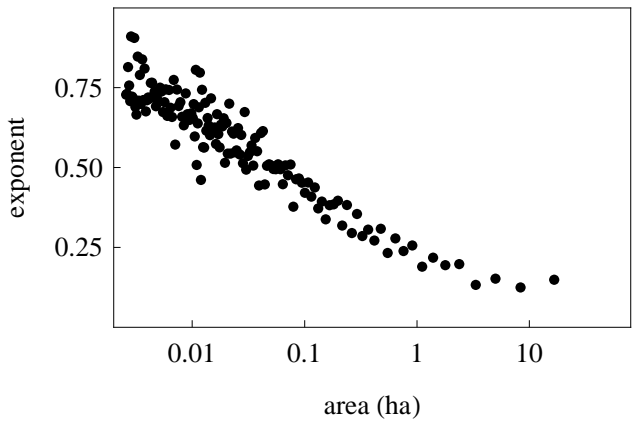}
  \end{center}
  \caption{Local slope (scaling exponent) of the curve plotted in Fig.~\ref{fig:bci_sar} for year 1995.}
  \label{fig:bci_exponent}
\end{figure}

For the BCI dataset the species-area scaling is not obvious at all, since the considered territory is 
quite small. As it is visible on Fig.~\ref{fig:bci_sar} the power-law trend appears only in the limit
of larger areas. The emergence of scaling becomes more visible if one plots the local exponent 
(local slope) of the curves plotted in Fig.~\ref{fig:bci_sar}. Considering the census year 
1995, on Fig.~\ref{fig:bci_exponent} we plot this variation. The plot suggests in the limit of 
larger territories (area larger than $1$ha) a stable power-law trend characterized by an exponent around 0.16 . 
As Fig.~\ref{fig:bci_sar} already suggests, similar results can be obtained for all census years, 
the stable scaling exponents varying between: 0.148 and 0.162. The obtained exponents are in 
agreement with species-area scaling results known in other macroecological systems \cite{harte_scaling_2000}. 

The $C(r)$ auto-correlation function, characterizing the spatial distribution of the 10 most abundant species was also calculated. For all of them, a power-law type decrease has been observed. As an example, on Fig.~\ref{fig:bci_corr} the auto-correlation function of the species that is most abundant among trees with a stem thicker than 10 cm, \emph{Trichilia tuberculata}, is visible (the spatial distribution of this species was illustrated on Fig.~\ref{fig:hybanthus_dens}). In Fig.~\ref{fig:bci_corr}a we plotted the results obtained on all trees belonging to this species and on Fig.~\ref{fig:bci_corr}b we plotted the results obtained for trees with diameter larger than 10 cm.  Both picture suggest a power-law like decrease for $C(r)$.

\section{A simple spatially explicit model}

There are numerous theoretical modeling efforts for understanding the relevant macroecological measures like the RSA distribution \cite{dewdney_dynamical_2000,chave_neutral_2004,chave_comparing_2002,enquist_modeling_2002,norris_neutral_2003,condit_census,condit_beta-diversity_2002,volkov_density_2005,volkov_patterns_2007,volkov_neutral_2003,volkov_organization_2004,volkov_neutral_2004,alonso_merits_2006,azaele_linear_2006,etienne_neutral_2007,pigolotti_stochastic_2004,zillio_spatial_2005}. Most of these models are however mean-field like approaches in the sense that the spatiality of the individuals is lost, and the only relevant quantity that characterizes the system is the number of individuals in each species. These models are usually successful for reproducing the right RSA curves. Our goal here is to follow the idea proposed by Zillio et al. \cite{zillio_spatial_2005} to go beyond the mean-field approach and to study a simple but still realistic, spatially explicit model. The tropical tree system, where the coordinates of individuals are fixed, is especially suited for such an approach.

The model considered by us is discrete both in time and space and it is inspired by the classical voter model \cite{holley_ergodic_1975}. Individuals are placed on a predefined lattice. Here, for the sake of simplicity a simple square lattice was considered. Each individual belongs to a given species and the state variable characterizing each lattice site codes these species. It is assumed that the lattice with sizes $L \times L$ is always completely filled up, corresponding to a constant ecological saturation of the territory. Thus the total number of individuals in the system is always $N= L \times L$. The number of possible states (species) is $W$, ($W \gg 1$).  The dynamics of the system is governed by two adjustable parameters: $p$ and $q$ ($0 < p, q \ll 1$).

In the beginning a randomly chosen Potts state (species) is assigned to each lattice site (individual). Starting from this random initial condition at each time moment a site is randomly chosen and its state is reconsidered. This process models the disappearance of one tree and birth of a new individual in the opened niche. The $p$ and $q$ parameters will govern the selection of the new species that occupies this free position. 
\begin{itemize}
  \item With probability $1-q-p$ the state of the chosen site is changed to one of its 8 closest neighbors. This is the most probable process, since both $p$ and $q$ are much smaller than one. It models the quite common phenomenon, in which a seed from a neighboring plant will successfully develop into a tree.
  \item With probability $q$ a randomly selected species is assigned to the chosen site. This species is selected with a uniform probability from the ensemble of possible species. If $W$ is big enough and $q$ much smaller than one, this process is suitable for modeling speciation, or immigration in the considered territory.
  \item With probability $p$ a species already existing in the lattice is assigned to the chosen site.  This species is selected by randomly choosing a site from the other $N-1$ available ones and identifying the species at that site. The process models diffusion of seeds on the considered territory, allowing seeds originating from far away individuals to reach the considered location.
\end{itemize}

\begin{figure}[ht]
  \begin{center}
    \includegraphics{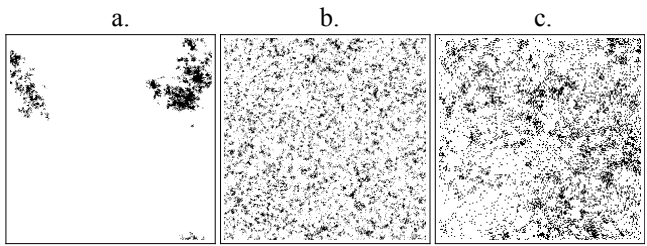}
  \end{center}
  \caption{Spatial distribution of the abundant species. Figure~a.\ shows simulation results with $p=0$, and figure~b.\ is for p=0.03. Figure~c.\ shows the spatial distribution of the abundant \emph{Trichilia tuberculata} species in the BCI-1995 census. The other simulation parameters are $L=500$, $q=1.5 \times 10^{-5}$.}
  \label{fig:sim_dens}
\end{figure}

In order to be in agreement with reality, the parameters $p$ and $q$ are selected as: $0 < q \ll p \ll 1$. For $p=0$ and $W \gg N$ we regain the analytically tractable model considered in \cite{zillio_spatial_2005}. Due to the combined short and long-range interactions the analytical treatment of the model seems quite difficult. The model is studied thus by Monte Carlo (MC) simulations. In order to minimize the effect of boundaries, periodic boundary conditions are used. One MC step is defined as $N$ updates of lattice sites. Simulations performed on several systems proved that in order to get realistic results for the final, statistically stable state both $q \ne 0$ and $p \ne 0$ is necessary. For $q = 0$ finally one of the species will colonize the whole territory. On the other hand, for $p = 0$ one would get unrealistically compact islands of species, in obvious contradiction with the fractal-like intermixing observed from the BCI dataset (see e.g.\ Fig.~\ref{fig:sim_dens}).

\section{MC simulation results}

\begin{figure}[ht]
  \begin{center}
    \includegraphics{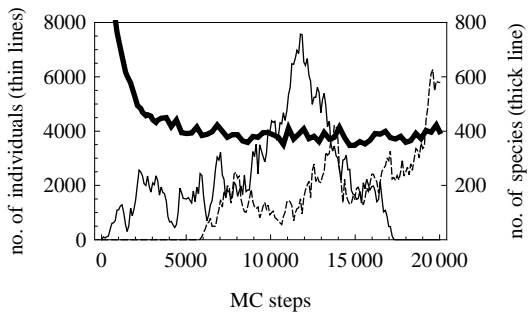}
  \end{center}
  \caption{Time evolution of the total number of species $W_s$ (thick continuous line) and the population of two selected species (continuous and dashed thin line). Simulations on a 
  $500 \times 500$ square lattice, $q=0.0001$, $p=0$.}
  \label{fig:time_evol}
\end{figure}

For $q \ne 0$ and $q \ll 1$ the model converges to a statistically stable state with several coexisting species. As an example, considering the $p=$0 simple case and choosing $L=500$, $q=0.0001$ on Fig.~\ref{fig:time_evol} we illustrate the time evolution of the total species number ($W_s$) in the system (thick continuous line).  After 5000 MC steps the $W_s$ species number reaches a stable limit and fluctuates around $W_s \approx 400$. On Fig.~\ref{fig:time_evol} we also show the time evolution of the population ($N_i$) for two selected species, one which appears at a later time moment and prevails, and one that appears quicker but gets extinct during the simulation (thin dashed and continuous lines, respectively).

In order to get a realistic spatial distribution for the abundant species, $p \ne 0$ is also necessary. 
For $p = 0$ all species will be distributed in island-like structures (Fig.~\ref{fig:sim_dens}a). 
Experimental results (Fig.~\ref{fig:sim_dens}c) show however a more homogeneous spatial distribution 
for all abundant species. By choosing a $0 < p \ll 1$ value, one will get results in better agreement with the experimental ones (Fig.~\ref{fig:sim_dens}b).

Several values for the $p$ and $q$ parameters ($0 < p,q \ll 1$) are appropriate for generating RSA and  species-area curves in reasonable agreement with the experimental results. Our goal here is to find those parameter 
values that will reproduce all the measured and previously discussed features of the BCI dataset.  First, we are looking for the $p$ and $q$ parameters that will lead to a statistically stable species number---around $300$---in the simulated area. Second, it is also desirable to get the relative size of the most abundant species (number of individuals in the most abundant species divided by $N$) of the same magnitude as in the experimental data. 
For the BCI dataset (census years 1982, 1985, 1990 and 1995) this relative size is: $0.169$, $0.17$, $0.165$ and $0.157$, respectively. Third, we are looking for a species-area scaling exponent in the $0.14$--$0.17$ range (Fig.~\ref{fig:bci_sar}) and an RSA probability density that shows a scaling with a $-1$ exponent and an obvious cutoff for abundant species (Fig.~\ref{fig:bci_rsa}c). 

\begin{figure}[ht]
  \begin{center}
    \includegraphics{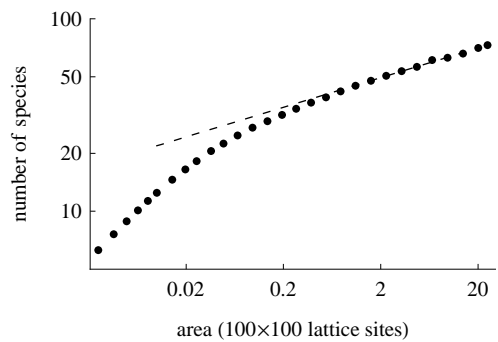}
  \end{center}
  \caption{Simulation results for the species-area curve (log-log scale) after 10000 MC steps. The power-law fit in the limit of large areas is drawn with a dashed line. The scaling exponent is $0.156$. The unit for area is chosen as 
$100 \times 100$ lattice sites. Simulation parameters: $L=500$, $p=0.03$ and $q=1.5 \times 10^{-5}$.}
  \label{fig:sim_sar}
\end{figure}

An extensive search by MC simulations in the $p$-$q$ parameter space suggested the $p=0.03$ and $q=1.5 \times 10^{-5}$ values. For these parameter values we got a stable species number around $300$ and the relative size of the most abundant species $0.14$. The species-area curve (Fig.~\ref{fig:sim_sar}) suggests a scaling exponent of $0.156$.

\begin{figure}[ht]
  \begin{center}
    \includegraphics[width=240pt]{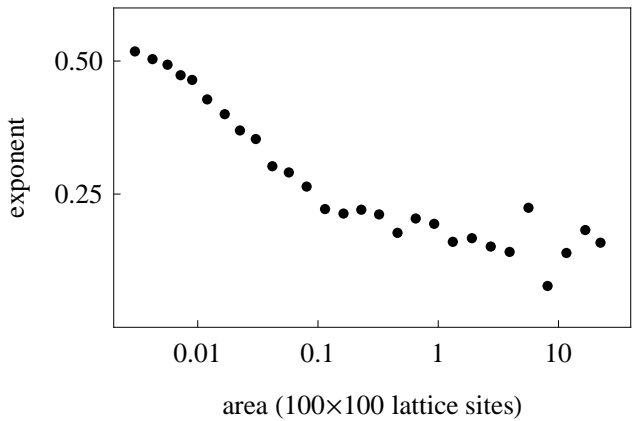}
  \end{center}
  \caption{Local slope (scaling exponent) of the curve plotted in Fig.~\ref{fig:sim_sar}.}
  \label{fig:sim_exponent}
\end{figure}

Moreover, finite-size effects in the species-area curve shape are also very similar with the ones
experimentally obtained. Plotting the local slope of the curve in Fig.~\ref{fig:sim_sar}, the trend presented in Fig.~\ref{fig:sim_exponent} is obtained. This trend can be compared with the one suggested by the BCI dataset (Fig.~\ref{fig:bci_exponent} for the 1995 census data).

\begin{figure}[ht]
  \begin{center}
    \includegraphics{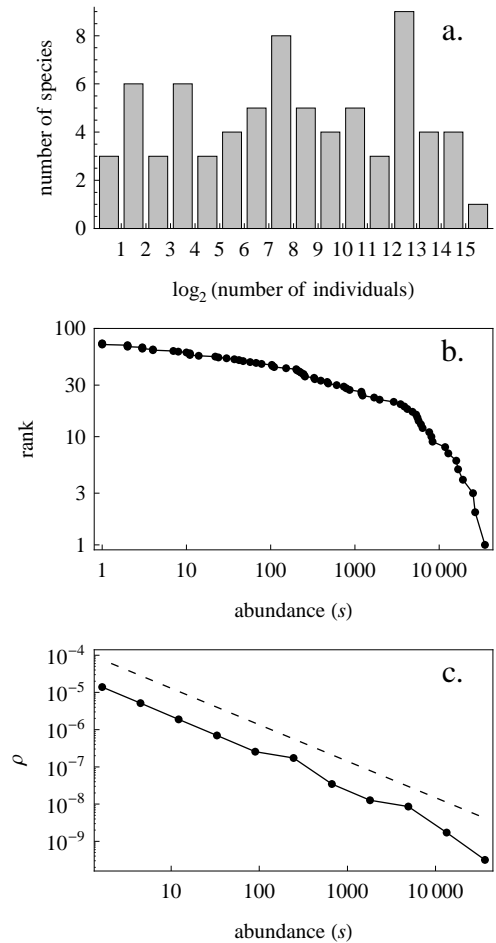}
  \end{center}
  \caption{Simulation results for the RSA curves after 10000 MC steps. Figure~a.\ shows the classical Preston plot,
figure~b. the rank-abundance curve and figure~c.\ the $\rho(s)$ probability-density function. The dashed line 
suggests a power-law with exponent $-1$. Simulation parameters: $L=500$, $p=0.03$ and $q=1.5 \times 10^{-5}$.}
  \label{fig:sim_rsa}
\end{figure}

The shape of the RSA curve is also in good agreement with the BCI results. Both the Preston plot 
(Fig.~\ref{fig:sim_rsa}a) and the probability density function (Fig.~\ref{fig:sim_rsa}b) resembles the experimentally observed ones (Fig.~\ref{fig:bci_rsa}). The $\rho(s)$ probability density has a scaling in the limit of rare species characterized by an exponent $-0.97$, close to the experimentally observed $-0.98$ value (Fig.~\ref{fig:bci_rsa}c).  

\begin{figure}[ht]
  \begin{center}
    \includegraphics{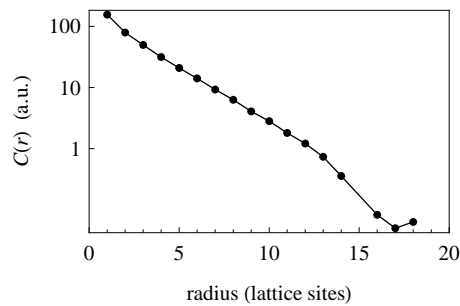}
  \end{center}
  \caption{The auto-correlation function of the most abundant species obtained after 10000 MC steps. Note that the scale for $C(r)$ is logarithmic. Simulation parameters are: $L=500$, $p=0.03$ and $q=1.5 \times 10^{-5}$.}
  \label{fig:sim_corr}
\end{figure}

Unfortunately, this spatially explicit model does not reproduce the observed power-law type decay (Fig.~\ref{fig:bci_corr}) of the auto-correlation function. The model suggest an exponential decay for the abundant species for all reasonable $p$ and $q$ parameter values. In particular, for the above discussed optimal $p$ and $q$ values, again an exponential decay is observed for the abundant species (Fig.~\ref{fig:sim_corr}).

\section{Discussion and Conclusions}

A simple spatially explicit model, which was first considered in \cite{zillio_spatial_2005}, was extended here for explaining the statistical properties of individuals and species in a tropical tree community. The model is defined on a square lattice and it is inspired by the classical voter model \cite{holley_ergodic_1975}. It describes in a realistic manner the spatiality of individuals, birth and death processes, diffusion of seeds and speciation events. Within the model the dynamics of species is governed by two adjustable parameters, $p$ and $q$. The parameter
$q$ defines the probability of speciation, or immigration events. The parameter $p$ describes the probability for the global seed diffusions within the considered territory.

The system was studied by large scale Monte Carlo simulations and the $p$ and $q$ parameters were adjusted for reproducing optimally all experimentally studied macroecological measures.  For reasonable values of $p$ and $q$ the model is successful in explaining the experimentally detected species number,  relative size of the most abundant species, the good species-area scaling and the shape of the RSA curves.  The model also generates a visually good spatial distribution of individuals within a species. The best model parameters were found as
$p=0.03$ and $q=1.5 \times 10^{-5}$. Beside all these successes, the model fails to reproduce the experimentally observed power-law decay for the auto-correlation function of the individuals within a species as a function of separation distance. As one would naturally expect from a lattice models where the local interactions are dominating ($p,q \ll 1$), simulations lead to an exponential decay.  This result suggests that in reality long-range interactions might be more important than it was considered in this simple approach. 

As final conclusion, it can be stated that in spite of its simplicity, the present model is successful in explaining the majority of the experimentally available macroecological measures. More complex models are needed however in order to fully understanding all aspects of spatial distribution of species in the tropical tree communities.

\begin{acknowledgments}
The present work was supported by a Sapientia KPI research grant.
\end{acknowledgments}

\end{document}